# *Albedo and laser threshold of a diffusive Raman gain medium*

A C SELDEN

Department of Physics
University of Zimbabwe
Mount Pleasant MP 167
HARARE Zimbabwe

*adrian_selden@yahoo.com*

ABSTRACT

The diffuse reflectance (albedo) and transmittance of a Raman random gain medium are calculated via semi-analytic two-stream equations with power-dependent coefficients. The results show good agreement with the experimental data for barium nitrate powder. Both the Raman albedo $A_R$ and Raman transmittance $T_R$ diverge at a critical gain $\gamma_c$, interpreted as the threshold for diffusive Raman laser generation. However, the ratio $T_R/A_R$ approaches a finite limiting value dependent on particle scattering albedo $\varpi$ and scattering asymmetry g. The dependence of the generation threshold on the scattering parameters is analysed and the feedback effect of Fresnel reflection at the gain boundaries evaluated. The addition of external mirrors, particularly at the pumped surface, significantly reduces the threshold gain.

Introduction

A number of non-linear optical effects have been observed in random laser media via the enhanced interaction arising from multiple scattering and gain, namely second *and higher* harmonic generation [1-3], anti-Stokes random lasing [4], up-conversion lasing [5] and surface plasmon enhanced Raman scattering [6] and random lasing [7]. Raman random lasing has been reported in SiC nanorods [8] and the Raman random laser threshold evaluated for a cloud of cold atoms [9]. Raman gain has been observed in optically pumped barium nitrate powder via enhanced reflectance and transmission gain of a Raman probe beam [10]. Unlike conventional random lasers, which function by optical excitation via the absorption bands, Raman random laser media require no intrinsic absorption, but operate by non-linear conversion of the pump light. As such they are ideally lossless, any residual absorption arising from impurities and surface contamination, thereby enabling the pump flux to reach much greater depths than in optically pumped random lasers. Raman gain observed in mono-crystalline barium nitrate powder has been modelled as a radiative transfer process in a scattering medium with non-uniform gain using Monte Carlo methods, the gain profile being determined from the observed variation of pump intensity with optical depth [10]. The pump radiation penetrates to a depth ~2 mm, equivalent to ~20 scattering lengths in a layer of randomly packed cubic crystals ~0.2-0.3 mm in size. A linear analysis of the propagation of light in powdered laser media has previously been made using the Kubelka-Munk two-flux equations with constant coefficients [11] and the dynamics of a 1D random laser modelled using the time-dependent diffusion equation [12]. Diffusion analysis has also been applied to model two-photon absorption in a random medium [13] and the distribution of second harmonic light in porous GaP [2]. Here we describe a semi-analytic two-stream model of diffusive Raman gain in a random medium, which calculates the diffuse reflected and transmitted radiation 'streams' directly, and gives good qualitative agreement with experimental data for *the* diffuse Raman reflectance and Raman transmission gain of barium nitrate powder [10]. Having validated the two-stream analysis, we apply it to determine the parametric dependence of the Raman albedo and Raman random laser threshold on the scattering characteristics of a random gain medium, with feedback provided by Fresnel and specular reflection of diffuse light at the gain boundaries [14, 15].

Two-stream analysis

Two-stream theory follows on integrating the radiative transfer equation over the forward and backward hemispheres, leading to a pair of coupled differential equations describing the spatial variation of the forward and backward radiative fluxes $F_+$, $F_-$

$$dF_+/d\tau = \gamma_{11} F_+ - \gamma_{12} F_- - S_+ \qquad (1a)$$

$$dF_-/d\tau = \gamma_{21} F_+ - \gamma_{22} F_- + S_- \qquad (1b)$$

where $\gamma_{mn}$ are transport coefficients, $S_+$, $S_-$ are source terms and $\tau$ is the optical depth [16]. The two-stream equations can be expressed in vector matrix form

$$d\mathbf{F}/d\tau = M\mathbf{F} + \mathbf{S} \qquad (2)$$

where $M(\gamma)$ is the transfer matrix, $\mathbf{F}(\tau)$ the vector flux and $\mathbf{S}(\tau)$ the vector source. This formalism is readily extended to higher order (multi-stream) analysis [16]. Selection of suitable source terms $S_+$, $S_-$ for photons scattered from the collimated probe beam enables calculation of the diffuse transmittance and hemispherical reflectance (albedo) of the scattering medium.

Transfer coefficients and rescaling

The standard Eddington approximation is used for the transport coefficients [16], modified to make the gain explicit [17]

$$\gamma_{11} = \gamma_{22} = \tfrac{1}{4}[7(1-\gamma) - \varpi(4+3g)] \qquad (3a)$$

$$\gamma_{12} = \gamma_{21} = -\tfrac{1}{4}[(1-\gamma) - \varpi(4-3g)] \qquad (3b)$$

with gain parameter $\gamma = \gamma_R \lambda_s$; the Raman gain coefficient $\gamma_R(\tau) = \Gamma_R F_p(\tau)$, where $\Gamma_R$ is the Raman gain parameter [10], $F_p(\tau)$ the diffuse pump flux, $\lambda_s$ the scattering length, $\varpi$ the particle scattering albedo and g the scattering asymmetry: $g = \tfrac{1}{2}\int \bar{p}(\mu_s)\mu_s d\mu_s$, where $\bar{p}(\mu_s)$ is the azimuthally averaged phase function, $\mu_s = \cos\theta_s$ and $\theta_s$ is the scattering angle (g $\Rightarrow$ 1 for forward scattering). The forward biased phase function is approximated by a simple two-term expression

$$\bar{p}(\mu_s) = 2\beta\delta|1-\mu_s| + (1-\beta)[1+3b_1\mu_s] \qquad (4)$$

the $\delta$-function representing the narrow diffraction peak containing a fraction $\beta$ of the total scattered energy, the linear term approximating the wide angle diffuse scattering distribution. The use of a $\delta$-function enables rescaling of the optical depth as $d\tau' = (1-\beta\varpi)d\tau$ and albedo as $\varpi' = \varpi(1-\beta)/(1-\varpi\beta)$ and reduces the phase function to the linear term [16]. For a Henyey-Greenstein phase function with asymmetry g we have $\beta = g^2$ and $b_1 = g/(1+g)$ [16]. The rescaled transfer coefficients are

$$\gamma'_{11} = \gamma'_{22} = \tfrac{1}{4}[7(1-\gamma')-\varpi'(4+3b_1)] \qquad (5a)$$

$$\gamma'_{12} = \gamma'_{21} = -\tfrac{1}{4}[(1-\gamma')-\varpi'(4-3b_1)] \qquad (5b)$$

where $\gamma' = \gamma/(1-\beta\varpi)$. The source terms for the diffuse Raman radiation scattered from the attenuated Raman probe flux $f_p(\tau)$ and their rescaled forms are

$$S_+ = \varpi\gamma_3 f_p(\tau) \Rightarrow S'_+ = \varpi'\gamma'_3 f_p(\tau') \qquad (6a)$$

$$S_- = \varpi(1-\gamma_3)f_p(\tau) \Rightarrow S'_- = \varpi'(1-\gamma'_3)f_p(\tau') \qquad (6b)$$

where $f_p(\tau) = f_p(0)\exp(-\tau/\mu_0)$, $\mu_0$ is the cosine of the angle between the inward surface normal and the incident beam ($\mu_0 = 1$ for normal incidence) and [16]

$$\gamma_3 = \tfrac{1}{4}[2 - 3g\mu_0] \Rightarrow \gamma'_3 = \tfrac{1}{4}[2 - 3b_1\mu_0] \qquad (7)$$

The rescaled two-stream equations are applied to calculate the diffuse reflectance and transmittance of the Raman random gain medium. The diffuse pump flux distribution $F_p(\tau)$ is obtained from the diffusion equation (see Appendix) and used to calculate the Raman gain profile $\gamma_R(\tau) = \Gamma_R F_p(\tau) = \gamma_0 h(\tau)$, where $h(\tau)$ is the depth dependence.

## Albedo equation

Defining $F_- = RF_+$ and substituting in eqns 1(a, b) yields the source-free albedo equation (with $S_+$, $S_- = 0$)

$$dR/d\tau = \gamma_{21} - (\gamma_{11} + \gamma_{22})R + \gamma_{12}R^2 \qquad (8)$$

where $R(\tau) = F_-(\tau)/F_+(\tau)$ is the diffuse reflectance function. The steady state solution is matched to the boundary reflectances $R(0) = R_0$ and $R(\tau_1) = 1/R_1$ [17]. Inserting the transfer coefficients defined in eqns 4(a, b) we have

$$dR/d\tau + 2\gamma_{11}(\tau)R = \gamma_{12}(\tau)(1+R^2) \qquad (9)$$

With the substitution $R = \tan \pi u/2$ for $u \in |0, 1|$, eqn (9) reduces to the compact form

$$du/d\tau + \gamma_{11}(\tau) \sin \pi u = \gamma_{12}(\tau) \qquad (10)$$

which can be used to determine the generation threshold for a particular configuration of the Raman gain medium with specified (reflecting) boundaries [17].

## Raman albedo

Two-stream calculations were carried out for collimated pump and probe beams incident on barium nitrate powder sandwiched between glass plates, for sample thicknesses in the range $L = 10\text{-}100\lambda_s$ [10]. The fluxes of amplified Raman radiation were determined by equating the hemispherical flux leaving a boundary with the specularly reflected incident flux: $F_-(0) = R_b F_+(0)$ and $F_+(\tau_0) = R_b F_-(\tau_0)$, where $R_b$ is the Fresnel reflection coefficient for diffuse radiation incident at a dielectric surface [14] and $\tau_0$ is the total optical depth. This condition is satisfied simultaneously at both boundaries for a specific value of the radiative flux, enabling the hemispherical reflectance (Raman albedo) and diffuse transmittance of the powder layer to be found. The Raman gains were obtained from the ratios of transmitted and reflected fluxes with and without the pump. The contribution of the diffuse pump flux reflected from the rear boundary was included in the gain profile [10]. For thin layers, multiple reflections at the boundaries are taken into account. For the barium nitrate powder samples used in the experiment, the inferred scattering parameters were $\lambda_s \approx 110$ μm and g ≈ 0.7, whence β ≈ 0.49 and $b_1$ ≈ 0.4. The Raman gain coefficient ranged from $\gamma_R$ = 0.5 cm$^{-1}$ to 2.2 cm$^{-1}$ [10]. A small but positive enhancement of the Raman albedo of the powder at the lowest Raman gain sets a lower limit on the particle scattering albedo viz. $\varpi$ ≥ 0.995. For high particle scattering albedoes, both pump and probe beams penetrate further into the diffusive medium, sampling relatively large gain volumes, the diffuse Raman radiation flux reaching its maximum value ~10-20 scattering lengths from the pumped surface (Fig 1). Thus quite modest Raman gain can result in significant amplification of the multiply scattered Raman radiation, as demonstrated by the enhanced reflectance and transmittance observed for barium nitrate powder [10]. Two-stream calculations of the Raman albedo $A_R$ and Raman transmission gain $T_R$ vs. L, using the experimentally determined depth dependence of the pump flux [10], generate profiles with closely similar characteristics to those observed (Fig 2), except for the quasi-exponential rise for the thinner layers (L ≤ 2 mm). However, the value of the gain parameter $\gamma_0$ required to fit the data points is lower than the experimental value as a result of the one-dimensional nature of the two-stream analysis, which neglects lateral diffusion of scattered light in the finite gain volume [10].

Raman laser threshold

Increasing the Raman gain by further increasing the incident pump intensity causes the diffusely reflected flux to rise rapidly, the Raman albedo $A_R$ and transmittance $T_R$ diverging as the Raman gain approaches a critical value $\gamma_c$, interpreted as the threshold gain parameter $\gamma_{th}$ for diffusive Raman laser generation. However, the ratio $T_R/A_R$ remains finite, reaching a limiting value $T_R/A_R|_{th}$ when $\gamma_0 = \gamma_{th}$ (Fig 3). The dependence of $T_R/A_R|_{th}$ on particle scattering albedo $\varpi$ and scattering asymmetry g is shown in Fig 4, the ratio increasing as $\varpi$ increases and decreasing as g increases. The laser threshold is significantly reduced with feedback provided by external mirrors, as observed for optically pumped random lasers [15, 18]. A 100% reflector at the rear boundary reduces the threshold for the thinner layers, the lowest threshold (~50% reduction) being reached when $L/\lambda_s$ = 20, but its effect disappears when $L/\lambda_s$ > 50 (Fig 5). A 100% reflector situated at the pumped surface has the maximum effect [19], halving the threshold for the thick powder layers (L > 50 mm). The lowest threshold occurs for the thinner layers with 100% reflectors at both boundaries, where both contribute to the feedback. For random gain media with high particle scattering albedoes i.e. low absorption loss, the diffuse Raman flux extends further from the pumped surface, reaching its maximum density some tens of scattering lengths from the pumped surface (Fig 6). The dependence of the threshold gain parameter $\gamma_{th}$ on particle scattering albedo $\varpi$ and scattering asymmetry g is shown in Fig 7, $\gamma_{th}$ decreasing as $\varpi$ increases i.e. as the loss per scattering diminishes; it shows a weaker dependence on g, slowly decreasing as g increases, except for $\varpi = 1$ (perfect scattering), when the boundary reflectance is dominant. The threshold gain parameter $\gamma_R \lambda_s$ is plotted against the diffuse attenuation parameter $\kappa_d \lambda_s$ for the pump flux in Fig 8, showing the monotonic increase in threshold gain with increased attenuation of the diffuse pump radiation. The dashed curves are power law fits to the data points of the form $y = a + bx^n$, where a, b, n are constants and $x = \kappa_d \lambda_s$. The exponent n lies in the range $1.76 \leq n \leq 1.96$, consistent with diffusion theory (n = 2). Thus the threshold gain parameter scales as $\gamma_{th} \lambda_s \propto (\lambda_s/\lambda_d)^2$ where $\lambda_d = 1/\kappa_d$ is the diffusion length.

Discussion

Given the simplicity of the two-stream analysis, it is satisfying that good qualitative agreement is found between the experimental and theoretical characteristics of the Raman albedo and transmission gain of barium nitrate powder. By adjusting the value of the single particle scattering albedo $\varpi$, the z-dependence of the pump flux profile is reproduced [10]; this value of $\varpi$ is then used to calculate the Raman gain profiles, with appropriate re-scaling to account for the narrow forward diffraction lobe of the particle scattering pattern (phase function). As a result of the high scattering albedo ($\varpi \geq 0.995$), both pump and probe light diffuse into the depths of the powder layer, such that for the thinner layers Fresnel reflection of the diffuse flux at both boundaries has to be taken into account. An obvious refinement of the analysis would be to include the effect of the radial gain profile corresponding to the gaussian profile of the incident pump beam, which diffuses laterally with increasing depth [10], and radial diffusion of the Raman radiation in the 'tear-drop' gain volume. In addition, we note that a gain threshold for diffusive Raman laser action in barium nitrate powder is predicted by the two-stream analysis, analogous with random laser generation via feedback scattering in random laser media [12]. The nature of laser generation in random media and its correct theoretical description is a topic of ongoing research [12, 22], such that a Raman random laser could best be demonstrated by experiment [8, 9]. However, the success of the two-stream model in analysing the Raman gain probe data for barium nitrate powder is encouraging and suggests it should be further tested against the experimental data for other non-linear media. The parametric dependence of the Raman random laser threshold is of particular interest as this tends to be model-dependent. The reduction in gain threshold predicted for powders with near-perfect scattering ($\varpi \approx 1$), and a significant reduction with reflecting boundaries, particularly the pumped surface, augur well for low threshold Raman powder lasers. The gain threshold also decreases for particles with higher scattering asymmetry i.e. larger particles, which could be tested by experiment [20]. The quasi-exponential increase in Raman albedo with depth observed for the thinner powder layers ($L \leq 2$ mm) is not explained by the two-stream model (nor by the Monte Carlo simulation) of diffusive Raman generation, suggesting that an alternative interpretation may be required for this regime, perhaps the excitation of internal resonances in individual crystallites [21] or excitation of specific laser modes in the random medium [22].

Conclusion

Two-stream analysis of the diffuse reflectance and transmittance of a Raman random gain medium gives good qualitative agreement with the experimental data for barium nitrate powder [10]. A gain threshold for diffusive Raman laser generation in a random medium is predicted on the basis of the analysis, which could be tested by experiment. The parametric dependence of the Raman laser threshold on particle scattering albedo $\varpi$, scattering asymmetry g and inverse diffusion length $\kappa_d$ has been determined and the reduction in gain threshold achieved by the addition of external reflectors evaluated. The low intrinsic absorption loss (ideally zero) associated with Raman random gain media allows order of magnitude increased penetration depths and interaction lengths compared with conventional optically pumped random lasers. As such, they allow stimulated Raman studies to be extended to mesoscopic random media, as is the case for second and higher harmonic generation [2, 3].

Appendix

The diffusion equation for the scalar flux (flux density) $\varphi$ follows on eliminating $F = F_+ - F_-$ from eqns. (1a), (1b) and writing $\varphi \approx 2(F_+ + F_-)$

$$\nabla^2 \varphi + S(\tau) = \kappa^2 \varphi \qquad (A1)$$

where $\nabla^2 = d^2/d\tau^2$, $\kappa^2 = \gamma_{11}^2 - \gamma_{21}^2$ and $S(\tau)$ is the source function [16, 17]. The coefficients $\gamma_{11}$, $\gamma_{21}$ defined by eqns 3(a), 3(b) above give

$$\kappa^2 = 3(1-\gamma-\varpi)(1-\gamma-\varpi g) \qquad (A2)$$

for the coefficient of $\varphi$ in the diffusion equation eqn (A1), which can be solved analytically for a planar incident beam (exponential source $S(\tau) = S_0 e^{-\tau}$) to yield the diffuse pump flux distribution (with $\gamma = 0$)

$$\varphi_p(\tau) = \varphi_0 [e^{-\kappa\tau} - \alpha e^{-\kappa(2\Lambda-\tau)} - \beta e^{-\tau}] \qquad (A3)$$

from which the Raman gain profile $\gamma_R(\tau) = \Gamma_R \varphi_p(\tau)$ can be derived ($\Lambda = L/\lambda_s$ and the coefficients $\alpha$, $\beta$ are determined from the boundary conditions). For nett gain $\gamma>1-\varpi$, $\kappa^2<0$; substituting $B^2 = -\kappa^2$ and setting $S(\tau) = 0$ yields the diffusive laser threshold equation [23] for the stimulated Raman flux $\varphi_R$

$$\nabla^2 \varphi_R + B^2(\tau)\varphi_R = 0 \qquad (A4)$$

where $B^2(\tau)$ is a diminishing function of optical depth, resulting from the decrease of Raman gain $\gamma_R(\tau)$ through attenuation of the diffuse pump flux $\varphi_p(\tau)$. When applied to diffusion in the powder layer, the flux $\varphi$ is extrapolated to zero a distance $z_e$ beyond the boundary, determined by the refractive index n ($z_e = 2.42$ for $n = 1.5$ [14, 24]). For anisotropic scattering, $z_e$ is expressed in terms of the transport length $\lambda_{tr} = \lambda_s/(1-\varpi g)$ [24]. For a perfectly reflecting boundary $z_e$ is infinite and the slope is set to zero: $d\varphi/d\tau|_{\tau=0} = 0$. Numerical solution of eqn (A4) with these boundary conditions gives generally good agreement with the two-stream analysis, except for minor differences in the calculated flux distributions near the boundaries. Diffusion theory has had some success in modelling powder lasers [12] and second harmonic generation in microporous GaP [2]

Fig. 1  Raman diffuse flux profiles: $F_+$ – flux approaching pumped surface, $F_-$ – flux leaving pumped surface, nett flux $F = F_+ - F_-$ ($F < 0$ when $F_- > F_+$), $f_p$ – diffuse probe flux. Curves plotted for $\varpi = 0.995$, $g = 0.7$.

Fig. 2  Raman albedo $A_R$ and Raman transmission gain $T_R$ of barium nitrate powder vs. thickness L of the powder layer for $\varpi = 0.995$, $g = 0.7$. The theoretical curves are matched to the experimental data points ■, ◆ by adjusting the gain parameter $\gamma_0$. The albedo saturates at smaller depths than the transmission gain, as observed.

Fig. 3  Dependence of Raman albedo $A_R$ and transmission gain $T_R$ on gain parameter $\gamma_0$, showing the divergence of $A_R$ and $T_R$ as $\gamma_0$ approaches the threshold value $\gamma_{th}$. The ratio $T_R/A_R$ remains finite, approaching a limiting value at threshold $T_R/A_R = 3.87…$ for the chosen parameters ($\varpi = 0.995$, $g = 0.7$).

Fig. 4  Threshold ratio $T_R/A_R|_{th}$ vs scattering albedo $\varpi$ ($g = 0.7$) Plane layer $L = 100\lambda_s$ Inset: $T_R/A_R|_{th}$ vs scattering asymmetry g ($\varpi = 0.995$).

Fig. 5  Threshold gain parameter $\gamma_R\lambda_s$ vs powder layer thickness $L/\lambda_s$ expressed in units of scattering length $\lambda_s$. The curves compare the influence of boundary reflectance on threshold gain: (a) powder layer between glass slides (b) 100% reflector at rear boundary (c) 100% reflectors at both boundaries.

Fig. 6  Diffuse Raman flux profiles at threshold vs single particle scattering albedo $\varpi$, showing the increasing penetration depth as $\varpi \Rightarrow 1$.

Fig. 7  Dependence of threshold gain parameter $\gamma_{th}$ on particle scattering albedo $\varpi$ and scattering asymmetry g.

Fig. 8  Threshold gain parameter $\gamma_R\lambda_s$ vs. diffuse attenuation parameter $\kappa_d\lambda_s$ for boundary reflectance R = 0, 0.57, 0.80, 1.00, showing the monotonic increase in threshold with increasing attenuation. Power law curves with exponents n in the range $1.8 \leq n \leq 2$ provide good fits to the calculated data points.

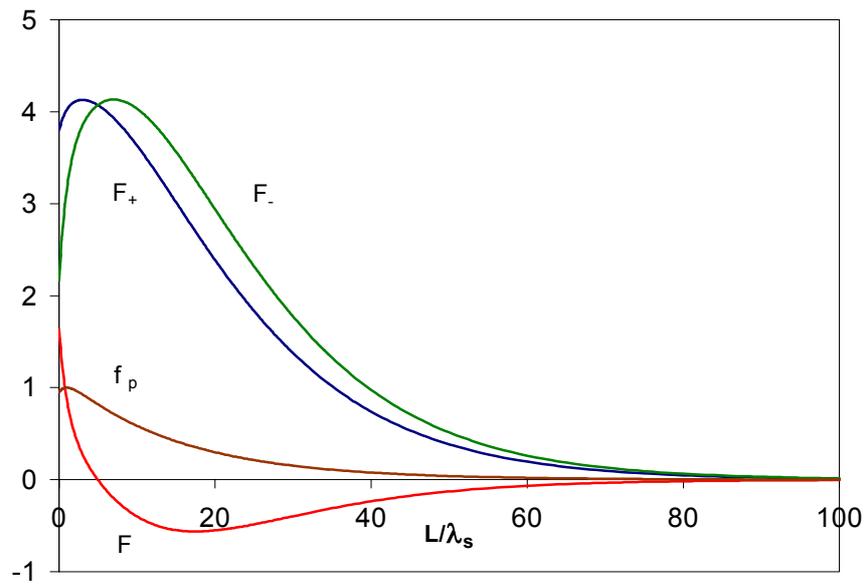

Fig 1

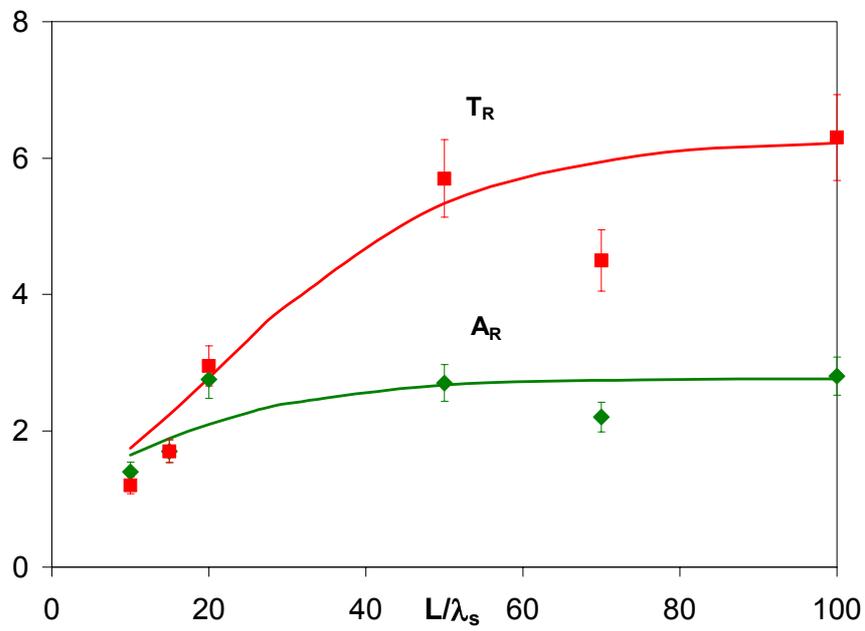

Fig 2

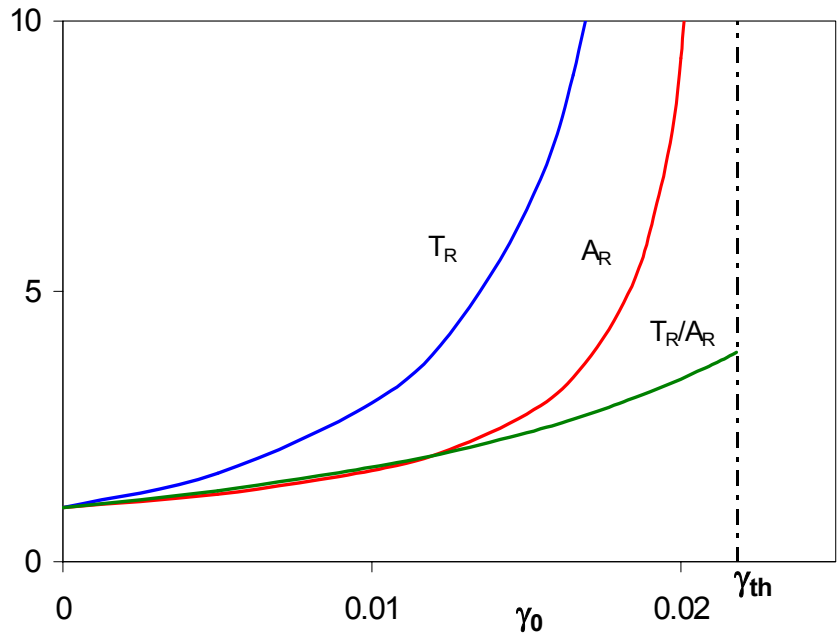

Fig 3

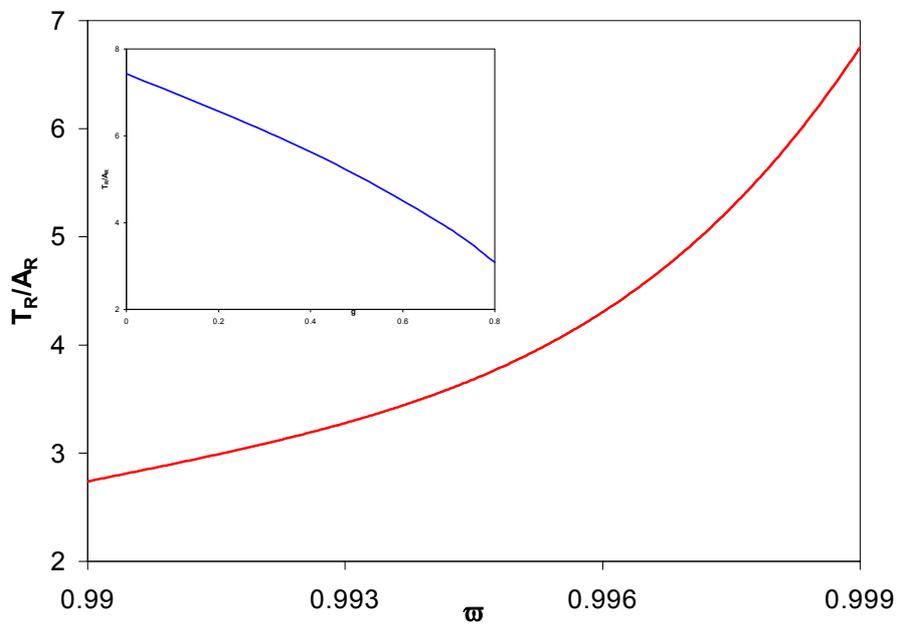

Fig 4

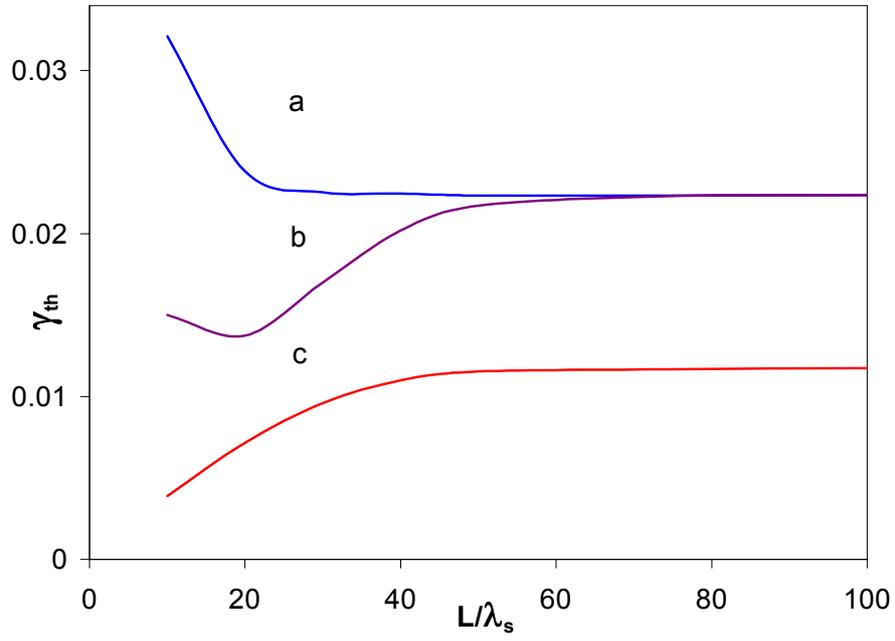

Fig 5

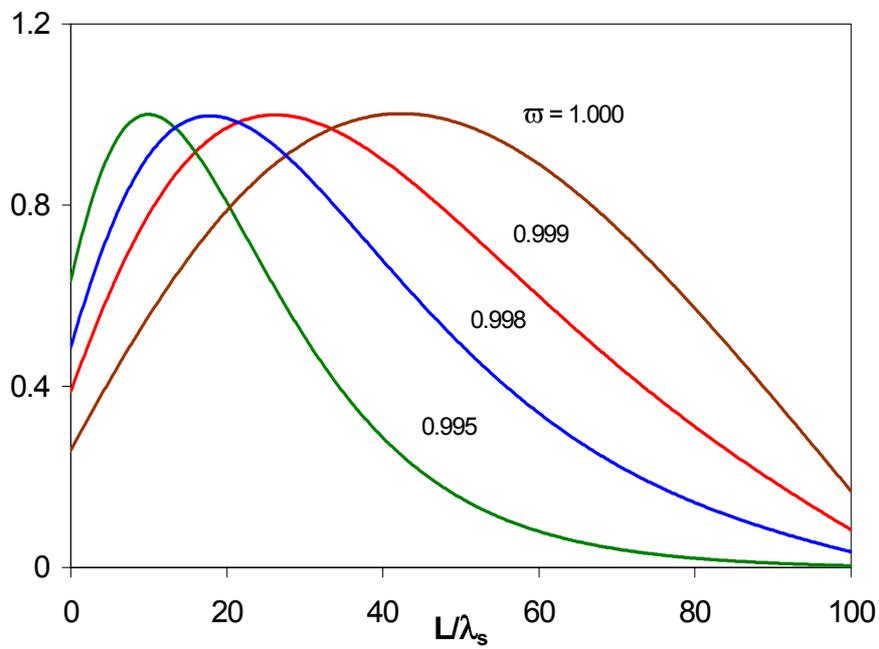

Fig 6

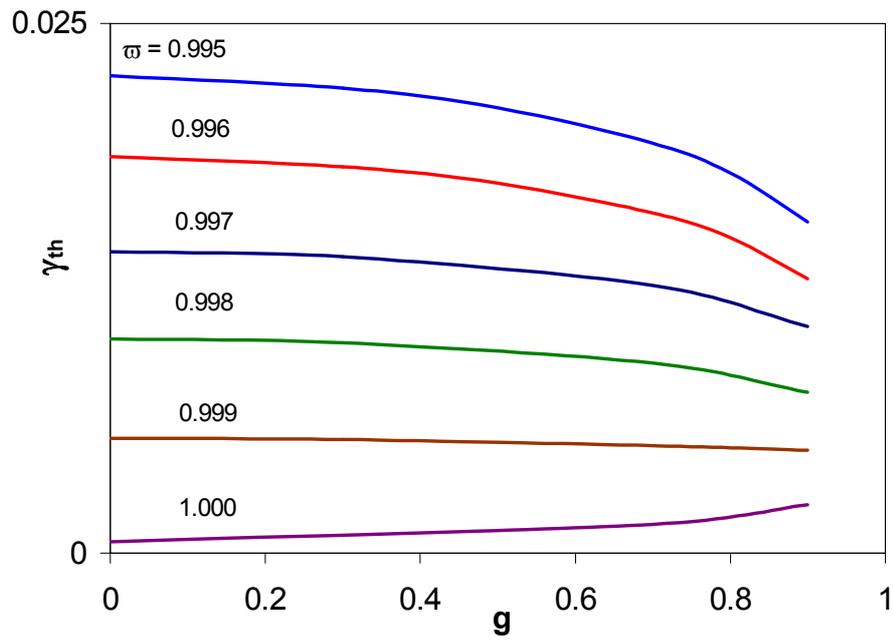

Fig 7

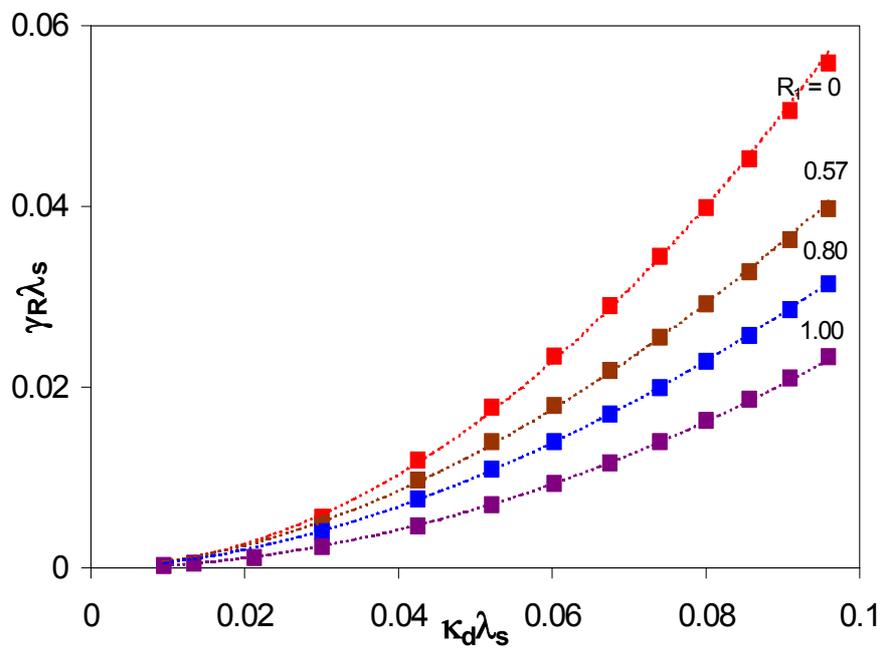

Fig 8